\documentclass[aps,preprint,showpacs,amsmath,amssymb]{revtex4}

\usepackage{amsmath,amssymb,amsfonts,bm,dcolumn,color,textcomp}
\usepackage[pdftex]{graphicx}

\newcommand{\vect}[1]{\mathbf{#1}}
\newcommand{\x}{\vect{x}}

\newcommand{\dx}{\vect{\delta x}}

\newcommand{\target}{\x^*}
\newcommand{\eps}{\epsilon}

\begin{document}

\thispagestyle{empty}
\begin{center}
{\bf\Large Controlling Complex Networks with \\ Compensatory Perturbations}
\end{center}

\vspace{0.4cm}

\centerline{Sean P. Cornelius$^1$, William L. Kath$^{2,3}$, and Adilson E. Motter$^{1,3}$}

\baselineskip 14pt

\begin{center}
{\it\small $^1$Department of Physics and Astronomy,Northwestern University, Evanston, IL 60208, USA}

{\it\small $^2$Department of Engineering Sciences and Applied Mathematics, Northwestern University, Evanston, IL 60208, USA}

{\it\small $^3$Northwestern Institute on Complex Systems, Northwestern University, Evanston, IL 60208, USA}
\end{center}

\vspace{0.8cm}

\noindent
{\bf 
The response of complex networks to perturbations is of utmost importance in
areas as diverse as ecosystem management,  emergency response, and cell
reprogramming. A fundamental property of networks is that the perturbation of
one node can affect other nodes, in a process that may cause the entire or
substantial part of the system to change behavior and possibly collapse. Recent
research in metabolic and food-web networks has demonstrated the concept that
network damage caused by external perturbations can often be mitigated or
reversed by the application of {\it compensatory} perturbations.  Compensatory
perturbations are constrained to be physically admissible and amenable to
implementation on the network. However, the systematic identification of
compensatory perturbations that conform to these constraints remains an open
problem. Here, we present a method to construct compensatory perturbations that
can control the fate of general networks under such constraints. Our approach
accounts for the full nonlinear behavior of real complex networks and can bring
the system to a desirable {\it target} state even when this state is not
directly accessible. Applications to genetic networks show that compensatory
perturbations are effective even when limited to a small fraction of all nodes
in the network and that they are  far more effective when limited to the
highest-degree nodes.  The approach is conceptually simple and computationally
efficient, making it suitable for the rescue, control, and reprogramming of
large complex networks in various domains.
}\\

\vspace{0.5cm}

\newpage
\noindent {\Large\bf Introduction} \\
\medskip

Complex systems such as  power grids, cellular networks, and food webs are often
modeled as networks of dynamical units. In such complex networks, a certain
incidence of perturbations and the consequent impairment  of the function of
individual units---whether genes, power stations, or species---are largely
unavoidable in realistic situations.  While local perturbations may rarely
disrupt a complex system, they can propagate through the network as the system
accommodates to a new equilibrium. This in turn often leads to system-wide
re-configurations that can manifest themselves as genetic diseases
\cite{motter_bioessays_2010,barabasi_review_2011}, power outages
\cite{Carreras_ieee_2004,Buldyrev_nature_2010}, extinction cascades
\cite{Pace_trends_1999,Scheffer_nature_2001}, traffic congestions
\cite{helbing_rmp_2001,Vespignani_science_2009}, and other forms of large-scale
failures \cite{May_nature_2008,Haldane_nature_2011}.

A fundamental characteristic of large complex networks, both natural and
man-made, is that they operate in a decentralized way. On the other hand, such
networks have either evolved or been engineered to inhabit stable states in
which they perform their functions efficiently. The existence of stable states
indicates that arbitrary initial conditions converge to a relatively small
number of persistent states. Such states are generally not unique and can change
in the presence of large perturbations. Because complex networks are
decentralized, upon perturbation the system can spontaneously go to a state that
is less efficient than others available.  For example, a damaged power grid 
undergoing a large blackout may still have other stable states in which no
blackout would occur, but the perturbed system does not converge to those
states.  We suggest that many large-scale failures are determined by the
convergence of the network to a ``bad" state rather than by the unavailability
of ``good" states.

Here we explore the hypothesis that one can design physically admissible
compensatory perturbations to direct a network to a desirable state even when it
would spontaneously go to an undesirable (``bad") state.  An important precedent
comes from the study of food-web networks, where perturbations caused by
invasive species or climate change can lead to the subsequent extinction of
numerous species. Recent research predicts that a significant fraction of these
extinctions can be prevented by the targeted suppression of specific species in
the system \cite{sagar_natcom_2011}.   Another precedent comes from the study of
metabolic networks of  single-cell organisms, where perturbations caused by
genetic or epigenetic defects can lead to nonviable strains, which are unable to
reproduce.  The  knockdown or knockout of specific genes has been predicted to
mitigate the consequence of such defects and often recover the ability of the
strains to grow~\cite{motter_msb_2008}. These findings have  analogues in power
grids, where perturbations caused by equipment malfunction  or operational
errors can lead to large blackouts, but the appropriate shedding of power can
substantially reduce subsequent failures
\cite{anghel_ieee_2007,Carreras_chaos_2002}.   Therefore, the concept underlying
our hypothesis is supported by recent research on physical
\cite{Carreras_chaos_2002,motter_prl_2004}, biological
\cite{motter_msb_2008,cornelius_pnas_2011}, and ecological networks
\cite{sagar_natcom_2011}. In these studies, the identification of compensatory
perturbations relied either on simple models of the real system
\cite{Carreras_chaos_2002,motter_prl_2004} or on using heuristic arguments
\cite{sagar_natcom_2011,motter_msb_2008}.  The question we pose is whether
compensatory perturbations can be systematically identified for a general
network of dynamical units.

Our solution to this problem is based on the observation that associated with
each desirable state there is a region of initial conditions that converge to
it---the so-called ``basin of attraction" of that state. Given a network that is
at (or will approach) an undesirable state, the conceptual problem is thus
reduced to the identification of  a perturbation in the state of the system that
can bring it to the attraction basin of the desired stable state (the {\it
target} state). Once there, the system will evolve spontaneously to the target. 
However, this perturbation must be physically admissible and is therefore
subject to constraints---in the examples above, certain genes can be
down-regulated but not over-expressed, the populations of certain species can
only be reduced, and changes in power flow are limited by capacity and
generation capability.  Under such constraints, the identification of a point
within the basin of attraction of the target state is a highly nontrivial task. 
 This is so because complex networks are high-dimensional dynamical systems and
it is generally impossible to identify the boundaries of a basin of attraction
in more than just a handful of dimensions. 

The approach introduced here overcomes this difficulty and is able to identify
compensatory perturbations in the absence of any {\it a priori} information
about the location of the attraction basin of the target state. This is done
systematically, without resorting to trial-and-error.  The approach is effective
even when only exploiting resources available in the network,  which usually
forbids bringing the system directly to the desired stable state.
This often
limits the search to perturbations that are locally deleterious, such as the
temporary inactivation of a node.  Interestingly, applications of our method show that such locally
deleterious perturbations can be tailored to lead to globally advantageous
effects. The approach is general and can be used to both control a network in
the imminence of a failure and to reprogram a network even in the absence of any
impending failure. \\ \\

\noindent {\Large\bf Results} \\
\medskip

\begin{figure}[!th]
\centering
\includegraphics[angle=0,width=12.5cm]{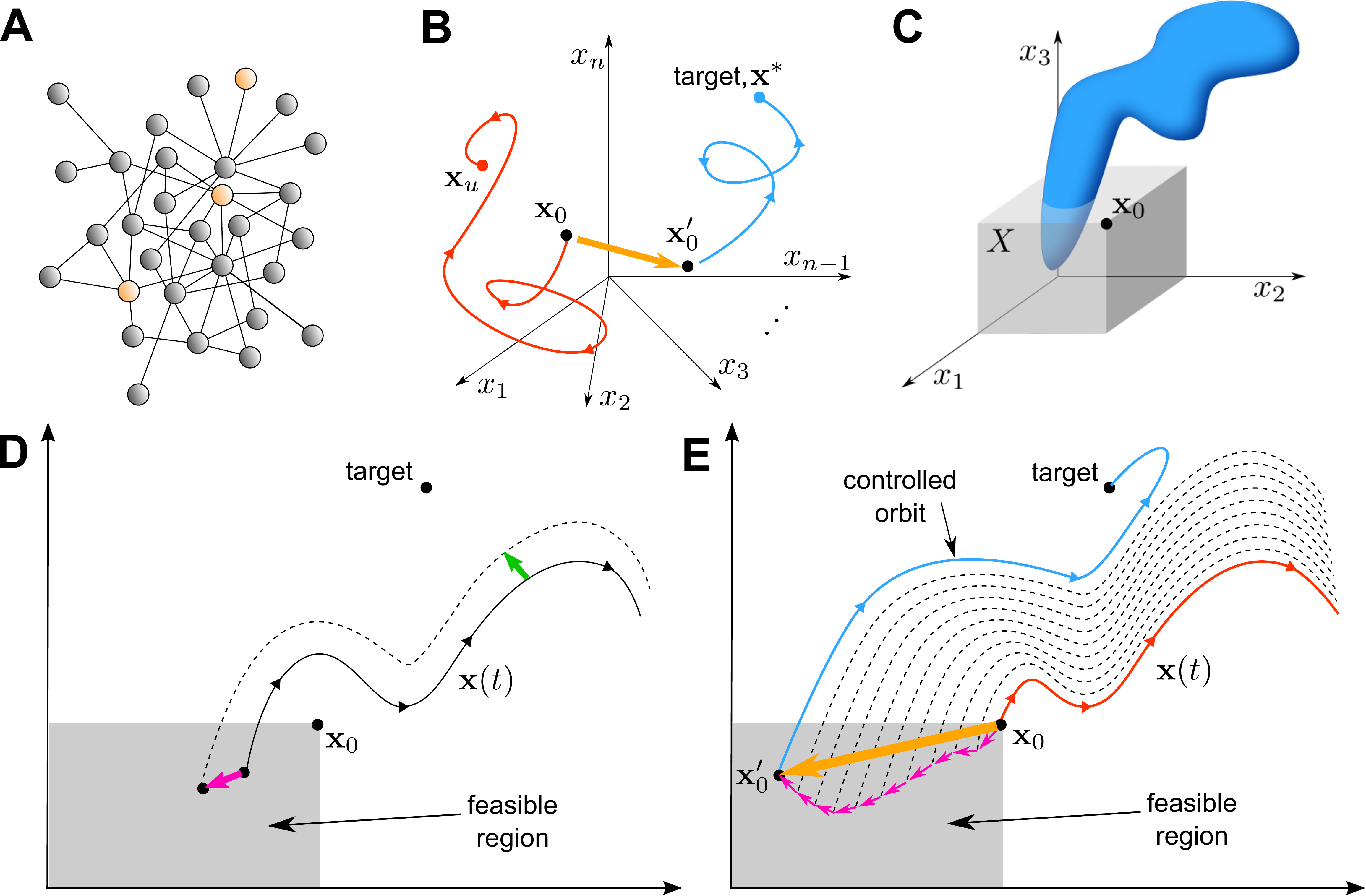}
\caption{\baselineskip 14pt
Schematic illustration of the network control problem.  (A) Network
portrait. The goal is to drive the network to a desired state by perturbing a
{\it control set} consisting of one or more nodes. (B) State space portrait. In
the absence of control, the network at an initial state $\x_0$ evolves to an
undesirable equilibrium  in the $n$-dimensional state space (red curve). By
perturbing this state (orange arrow), the network reaches a new state  $\x_0'$
that evolves to the desired target state (blue curve). (C) Constraints. In
general, there will be constraints on the types of compensatory perturbations
that one can make. In this example, one can only perturb three out of $n$
dimensions (equality constraints), which we assume to correspond to a thee-node
control set, and the dynamical variable along each of these three dimensions can
only be reduced  (inequality constraints). This results in a set of eligible
perturbations, which in this case forms a cube within the three-dimensional 
subspace of the control set.  The network is controllable if and only if the
corresponding slice of the target's basin of attraction (blue volume) intersects
this region of eligible perturbations (grey volume).  (D,  E)  Iterative
construction of compensatory perturbations. (D) A perturbation to the current
initial condition (magenta arrow) corresponds to a perturbation of the resulting
orbit (green arrow) at the point of closest approach to the target.  At every
step, we seek to identify a perturbation that brings this point, and hence the
orbit, closer to the target. (E) This process generates orbits that approach
increasingly closer to the target (dashed curves), and is repeated until a
perturbed state  $\x_0'$  is identified that evolves to the target.  The
resulting compensatory perturbation  $\x_0'-\x_0$ (orange arrow) brings the
system to the attraction basin of the target without any {\it a priori}
information about its location, and allows directing the network to a state that
is not directly accessible by any eligible perturbation. } 
\label{fig1}
\end{figure}

Figure \ref{fig1}A-C illustrates the problem that we intend to address. The
dynamics of a network is best studied in the state space, where we can follow
the time evolution of individual trajectories and characterize the stable states
of the whole system.  Figure \ref{fig1}A represents a network that would
spontaneously go to an undesirable state, and that we would like to bring to a
desired stable state by perturbing at most three of its nodes (highlighted). 
Using  $\x = (x_1, ..., x_n)$ to represent the dimensions of the state space, 
Fig.\ \ref{fig1}B shows how this perturbation applied at a certain time $t_0$,
changing the state of the system from $\x_0$ to $\x_0'$, would lead to an orbit
that asymptotically goes to the target state.  As an additional constraint,
assume that the activity of the nodes can only be reduced (not increased) by
this perturbation. Then, in the subspace corresponding to the nodes that can be
perturbed, the set of points $X$ that can be reached by eligible compensatory
perturbations forms a cubic region, as shown in  Fig.\ \ref{fig1}C.  The target
state itself is outside this region, meaning that it cannot be directly reached
by {\it any} eligible perturbation. However, the basin of attraction of the
target state may have points inside the region of eligible perturbations (Fig.\
\ref{fig1}B), in which case the target state can be reached through
perturbations that bring the system to one of these points; once there, the
system will spontaneously evolve towards the target state. This scenario is
representative of the conditions under which it is important to identify
compensatory perturbations, and leads to a very clear conclusion: a compensatory
perturbation exists if and only if the region formed by eligible compensatory
perturbations overlaps with the basin of attraction of the target.   

However, there is no general method to identify basins of attraction (or this
possible overlap) in high-dimensional state spaces typical of complex networks.
Indeed, despite significant advances, numerical simulations are computationally
prohibitive  and analytical methods, such as those based on  Lyapunov stability
theorems, offer only rather conservative estimates and are not yet sufficiently
developed to be used in this context  \cite{Genesio1985,Kaslik2005}.
Accordingly, our approach does not assume any information about the location of
the attraction basins and addresses a problem that, as further explained below,
cannot be solved by existing methods from control theory. \\

\medskip
\noindent {\large\bf Control procedure for networks}
\bigskip

The dynamics of a complex network can often be represented by a set of coupled ordinary differential equations. We thus consider an $N$-node network whose  $n$-dimensional dynamical state $\x$ is governed by 
\begin{equation} \label{system} 
\frac{d\x}{dt} = \vect{F}(\x).
\end{equation}
The example scenario we envision is the one in which this network has been
perturbed at a time prior to $t_0$, driving it to a state $\x_0=\x(t_0)$ in the
attraction basin $\Omega(\x_u)$ of an undesirable state $\x_u$. The compensatory
perturbation that we seek to identify consists of a judiciously-chosen
perturbation $\x_0 \rightarrow \x_0'$ to be implemented at time $t_0$, where $
\x_0'$ belongs to the basin of attraction $\Omega(\target)$ of a desirable state
 $\target$.  For simplicity, we assume that $\x_u$ and $\target$ are fixed
points, although the approach we develop extends to other types of attractors.
In the absence of any constraints it is always possible to perturb $\x_0$ such
that  $\x_0' \equiv \target$. However, as illustrated above, usually only 
constrained compensatory perturbations are allowed in real networks. These
constraints encode practical considerations and  often take the form of
mandating no modification to certain nodes, while limiting the direction and
extent of the changes in others. The latter is a consequence of the relative
ease of removing versus adding resources to the system. We thus assume that the
constraints on the eligible perturbations can be represented  by the vectorial
expressions 
\begin{equation} \label{constraints}
\mathbf{g}(\x_0, \x_0') \leq {\bf 0} \,  \mbox{ and } \, \mathbf{h}(\x_0, \x_0') = {\bf 0},
\end{equation}
where both the equality and the inequality are assumed to apply to each component.  
In a food web, for example, admissible compensatory perturbations may only reduce the populations of
certain unendangered species, while leaving those of the endangered ones
unchanged. For concreteness, we focus on constraints of the form $x_{0j}'\le
x_{0j}$ for accessible nodes, and $x_{0j}'= x_{0j}$ for the other nodes, 
although the approach we develop accommodates general, nonlinear constraints. 

We propose to construct compensatory perturbations iteratively from small
perturbations, as shown in Fig.\ \ref{fig1}D-E. Given a dynamical system in the
form (\ref{system}) and an initial state $\x_0$ at time $t_0$, a small
perturbation $\delta \x_0$  evolves to  $\delta \x(t)={\bf M}(\x_0, t)\cdot \delta \x_0$ at time $t$. 
The matrix ${\bf M}(\x_0, t)$ is the solution of the
variational equation $d{\bf M}/dt= D{\mathbf F}(\x)\cdot {\bf M}$ subject to the
initial condition ${\bf M}(\x_0, 0)={\bf 1}$, and is generally invertible
\cite{ott02}. Given the time $t_c$ of closest approach of the perturbed orbit to
the target $\target$, we can in principle use the inverse transformation,
\begin{equation}
\delta \x_0={\bf M}^{-1}(\x_0, t_c)\cdot \delta \x(t_c),
\label{variational}
\end{equation}
to determine the perturbation $\delta \x_0$ to the initial condition $\x_0$
that, among the admissible perturbations satisfying $|\delta
\x_0|\le\epsilon_1$, will render $\x(t_c) + \delta\x(t_c)$ closest closest to
$\target$ (Fig.\ \ref{fig1}D). This is a good approximation for small
$\epsilon_1$, and hence for small changes in the initial conditions; the error
can in fact be calculated from the matrix of second derivatives and verified
numerically by direct integration. Large perturbations can then be built up by
iterating the process: every time $\delta \x_0$ is calculated, the current
initial state, $\x_0'$, is updated to $\x_0'+\delta \x_0$, and a new $\delta
\x_0$ is calculated starting from the new initial condition (Fig.\ \ref{fig1}E).
 
The problem of identifying a perturbation $\delta \x_0$ that incrementally moves
the orbit toward the target under the given constraints is then cast as a
constrained optimization problem (see \emph{Methods}).  Once
found, the optimal incremental perturbation is applied to the current initial
condition,  $\x_0' \rightarrow \x_0' + \dx_0$,  giving the initial state for the
next iteration.  At this point, we test whether the new state lies in the
target's  basin of attraction by integrating the system (\ref{system})  over a
long time $\tau$.  If the resulting orbit reaches a small ball of radius
$\kappa$ around the target,  we declare  success and terminate the procedure.
Otherwise, we  identify the closest approach point in a time window $t_0 \leq t
\leq t_0+ T$, where $T$ is a time limit determined beforehand, and repeat the
procedure. Now, it may be the case that no compensatory perturbation can be 
found, e.g., if the feasible region $X$ does not intersect the target basin
$\Omega(\target)$.  To account for this, we automatically terminate our search
if the  system is not controlled within a fixed number of  iterations (see
\emph{Methods}). 

\begin{figure}[!th]
\centering
\includegraphics[width=15cm]{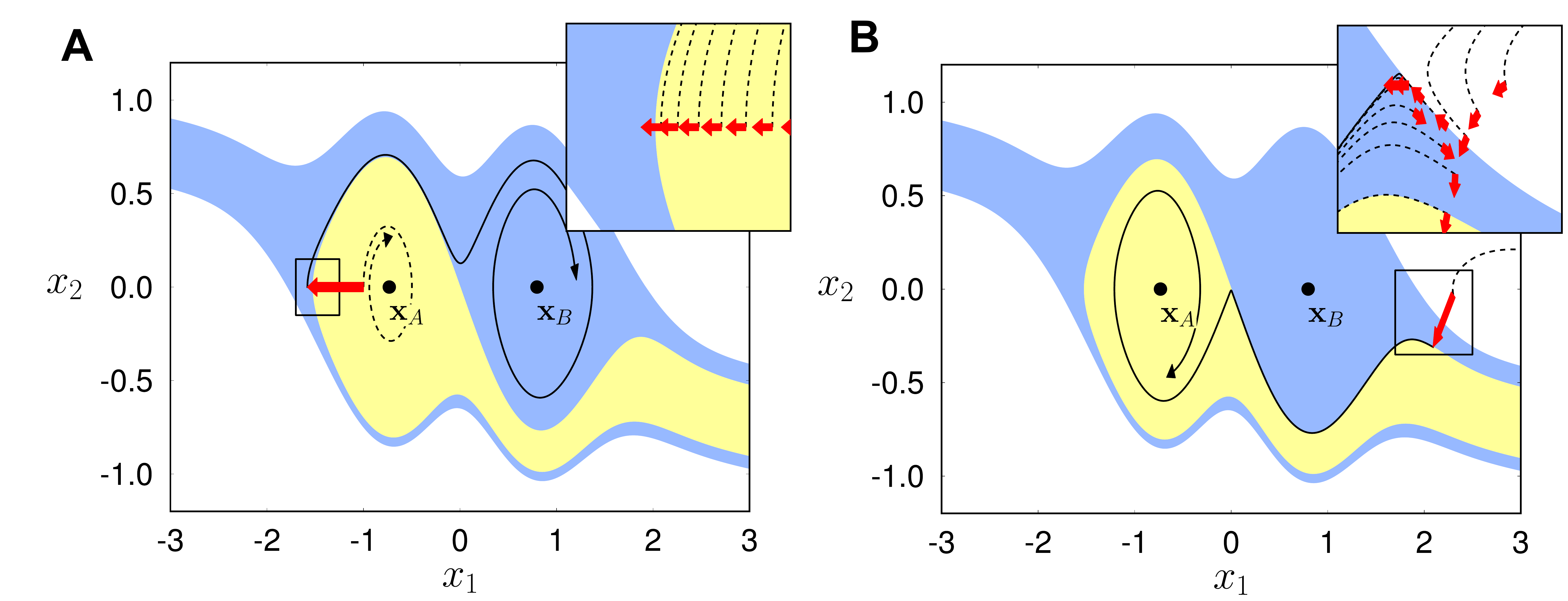}
\caption{\baselineskip 14pt
Illustration of the control process in two dimensions. Yellow and blue
represent the basins of attraction of the stable states $\x_A$ and $\x_B$,
respectively,  while white corresponds to unbounded orbits. (A and B) Iterative
construction of the perturbation for an initial state in the basin of $\x_A$ 
with $\x_B$ as a target (A), and for an initial state on the right side of both
basins with $\x_A$ as a target (B).  Dashed and continuous lines indicate the
original and controlled orbits, respectively. Red arrows indicate the full
compensatory perturbations. Individual iterations of the process are shown in
the insets (for clarity, not all iterations are included). } 
\label{fig2}
\end{figure}

Before considering the application to networks, we first illustrate our method
using an example in two dimensions, where the basins of attraction (and hence
the possible compensatory perturbations) can be easily calculated and
visualized.  Figure \ref{fig2} shows the state space of the system, which has
two stable states: $\x_A$ on the left and $\x_B$ on the right. The system is
defined by the potential $U(x_1)= \exp (-\gamma x_1^2)(bx_1^2+cx_1^3+dx_1^4)$
and frictional dissipation $\eta$,  where $\gamma=1$, $b=-1$, $c=-0.1$, $d=0.5$,
$\eta=0.1$, and $x_2=\dot{x}_1$. The method is illustrated for two different
initial states under the constraint that admissible perturbations have to
satisfy  $\x_0'\le \x_0$, i.e., one cannot increase  either variable. For the
initial state in the basin of state $\x_A$ (Fig.\ \ref{fig2}A), no admissible
perturbation exists that can bring the system directly to the target $\x_B$, on
the right, since that would require increasing $x_1$. However, our iterative
procedure builds an admissible perturbation vector that shifts the state of the
system to a branch of the basin $\Omega(\x_B)$ lying on the left of that point.
Then, from that instant on the autonomous evolution of the system will govern
the trajectory's approach to the target $\x_B$, on the right. This example
illustrates how compensatory perturbations that move in a direction away from
the  target---the only ones available under the given constraints---can be
effective in controlling the system, and how they are identified by our method.
The other example shown illustrates a case in which the perturbation to an
initial state on the right crosses an intermediate basin, of $\x_B$, before it
can reach the basin of the target, $\x_A$ (Fig.\ \ref{fig2}B). The linear
approximation fails at the crossing point, but convergence is nonetheless assured by the
constraints imposed on $\delta\x_0$ (see \emph{Methods}). \\

\medskip
\noindent {\large\bf Control of genetic networks}
\bigskip

\begin{figure}[th]
\centering
\includegraphics[width=12.5cm]{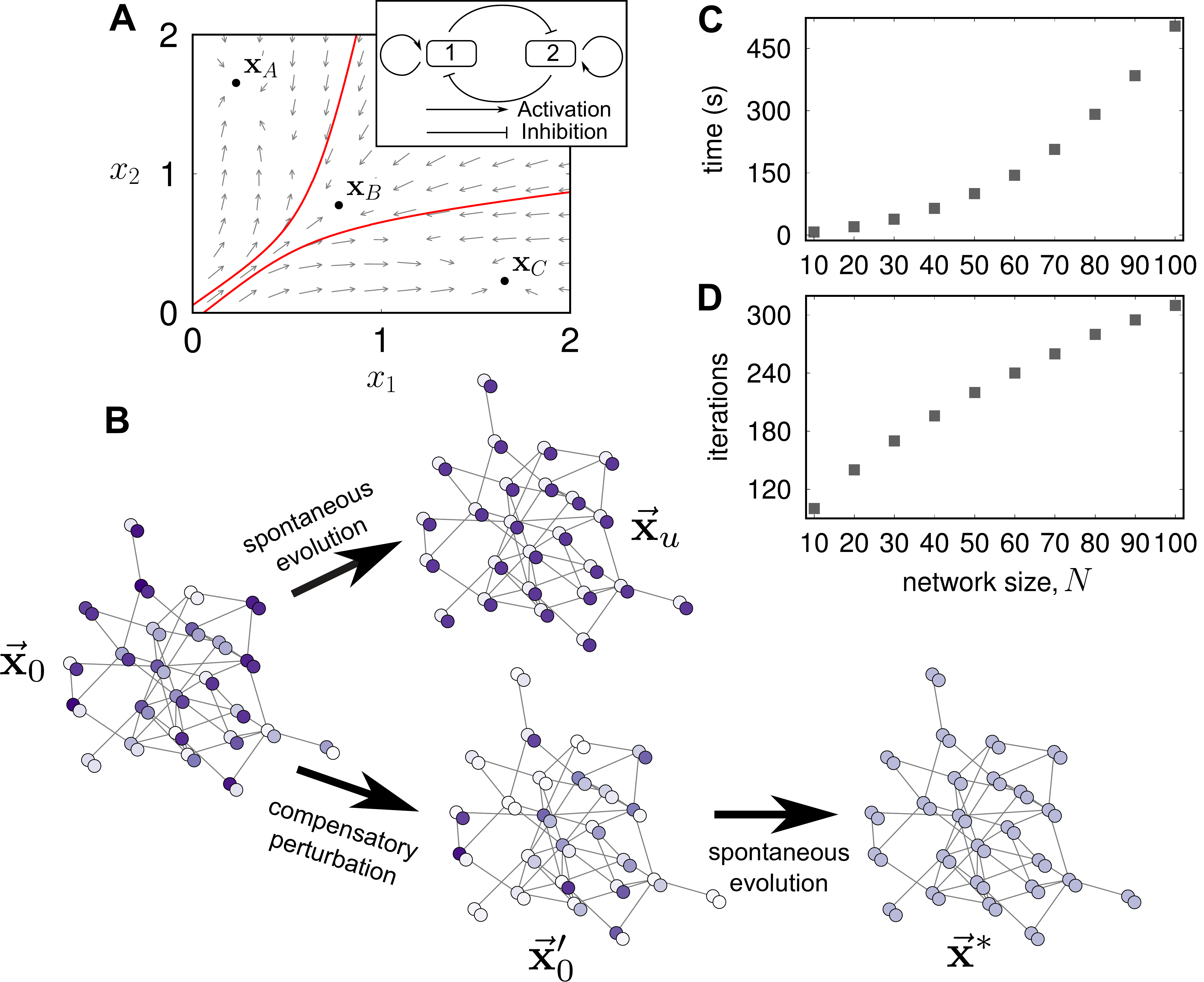}
\caption{\baselineskip 14pt
Control of genetic networks. 
(A) State space of the two-gene subnetwork represented in the inset, where the
red curves mark the boundaries between the basins of $\x_A$, $\x_B$, and $\x_C$,
and the arrows indicate the local vector field.  (B) Illustration of
compensatory perturbation on complex genetic networks, where each node is a copy
of the two-gene system. We are given an initial network state $\vec{\x}_0$
representing the expression levels of the  $N$ gene pairs (color coded), and
this state evolves to a stable state of the network $\vec{\x}_u$ (top path). 
The goal is to knockdown one or more genes to reach a new state $\vec{\x}_0'$
that instead evolves to a  target stable state $\vec{\x}^*\neq \vec{\x}_u$
(bottom path). (C and D) Average computation time (C) and average number of
iterations (D) required to control networks of $N$ nodes  with initial state
$\vec{\x}_0 = \vec{\x}_A$ and target $\vec{\x}^* = \vec{\x}_B$, demonstrating
the good scalability of the algorithm. Each point represents an average over
$1,000$ independent realizations of homogeneous networks.} 
\label{fig3}
\end{figure}

We now turn to compensatory perturbations in complex networks. We focus on
networks of diffusively coupled units, a case that has received much attention
in the study of spontaneous synchronization \cite{pecora_prl_2008}.  We take as
a base system the genetic regulatory subnetwork shown in Fig.\  \ref{fig3}A
(inset), consisting of two genes wired in a circuit. The state of the system is
determined by the expression levels of the genes, represented by the variables
$x_1, x_2 \geq 0$. The associated dynamics obeys
\begin{align} 
\frac{dx_1}{dt} &= a_1 \frac{x_1^m}{x_1^m + S^m} + b_1 \frac{S^m}{x_2^m + S^m} - k_1 x_1 + f_1,  \label{twogene}  \\
\frac{dx_2}{dt} &= a_2 \frac{x_2^m}{x_2^m + S^m} + b_2 \frac{S^m}{x_1^m + S^m} - k_2 x_2 + f_2,  \label{twogene2}
\end{align}
where the first two terms for each gene capture the self-excitatory and mutually
inhibitory interactions represented in Fig.\ \ref{fig3}A, respectively, while
the final two terms represent linear decay and a basal activation rate of the
associated gene's expression. Models of this form have been used extensively to describe
the transition between progenitor stem cells and differentiated cells
\cite{Roeder_jtb_2006,Huang_DevBiol_2007,huang_biophysj_2009}. 
For a wide range
of parameters, this system exhibits three stable states: a state ($\x_B$)
characterized by comparable expression of both genes (the stem cell state), 
and two states  ($\x_A$ and $\x_C$) characterized by the dominant expression of one of the genes
(differentiated cell states). 
Figure \ref{fig3} and subsequent results correspond to the symmetric choice of
parameters $a_{1,2}=0.5$, $b_{1,2}=1$, $k_{1,2}=1$, $f_{1,2}=0.2$, $S = 0.5$,
and $m=4$. We assume that compensatory perturbations are limited to gene
down-regulations, i.e., $\x_0'\le \x_0$. Admissible compensatory perturbations do not
exist (and hence cannot be identified) from some other initial states close to
the axis in the basin of $\x_{A,C}$  having $\x_{C,A}$  as target, indicating
that the expression of that gene is too low to be rescued by compensatory
perturbations. But $\x_B$ can be reached even in these cases, and thus so can
$\x_{A,C}$ if we let the system evolve after a perturbation into $\Omega(\x_B)$
and then perturb it again into $\Omega(\x_{C,A})$, meaning that we need the
conversion from the differentiated state to the stem cell state before going to
the other differentiated state. 

We construct large genetic networks by coupling
multiple copies of the two-gene system described above. Such networks may
represent cells in  a tissue or culture coupled by means of factors exchanged
through their microenvironment or medium. Specifically, we assume that  each
copy of this genetic system can be treated as a node of the larger network. The
dynamics of a network consisting of $N$ such systems is then governed  by
\begin{equation} \label{syncdynamics}
\frac{d\x_i}{dt} = \vect{f}(\x_i) + \frac{\varepsilon}{d_i} \sum_j A_{ij} [\x_j  - \x_i],
\end{equation}
where $\dot{\x}_i = \vect{f}(\x_i)$ is the vectorial form of the dynamics of
node $i$ as described by Eqs.\  (\ref{twogene})-(\ref{twogene2}), the parameter
$\varepsilon>0$ is the overall coupling strength, and $d_i$ is the degree
(number of connections) of node $i$. The structure of the network itself is
encoded in the adjacency matrix  $A = (A_{ij})$.  We focus on randomly generated
networks with both uniform and preferential attachment rules (\emph{Methods}),  
which serve as models for homogeneous and heterogeneous degree
distributions, respectively.  We use $\vec{\x}=(\x_i)$ to denote the state of
the network, with $\vec{\x}_A$,  $\vec{\x}_B$,  and $\vec{\x}_C$ denoting the
states in which all nodes are at state $\x_A$,  $\x_B$,  and $\x_C$,
respectively.  The states $\vec{\x}_A$, $\vec{\x}_B$,  and $\vec{\x}_C$ are
fixed points of the full network dynamics in the $2N$-dimensional state space
and, by arguments of structural stability, we can conclude they are also stable
and have qualitatively similar basins of attraction along the coordinate planes
$\x_i$ if the coupling strength $\varepsilon$ is weak. While we focus on these
three states,  it follows from the same arguments that in this regime there are
$3^N\! - 3$ other  stable states in the network. Under such conditions,
compensatory perturbations between $\vec{\x}_A$,  $\vec{\x}_B$,  and
$\vec{\x}_C$ are guaranteed to exist, and hence this class of networks can also
serve as a benchmark to test the effectiveness and efficiency of our method in
finding compensatory perturbations in systems with a large number of nodes.  

A general compensatory perturbation in this network is illustrated  in Fig.\
\ref{fig3}B, where different intensities indicate different node states. Applied
to the initial state  $\vec{\x}_A$ and target  $\vec{\x}_B$ for 
$\varepsilon = 0.05$, the method is found to be effective in $100\%$ of the cases for the
$10,000$ networks tested, with $N$ ranging from $10$ to $100$. Moreover,  the
computation time and number of iterations for these tests confirm that our
method is also computationally efficient for large networks.  Computation time
grows approximately quadratically with $N$ (Fig.\ \ref{fig3}C), as expected
since each iteration requires the integration of  $O(N^2)$ equations.   The
number of iterations grows as the square root of $N$ (Fig.\ \ref{fig3}D), in
agreement with the $\sqrt{N}$ scaling of  $\lvert \vec{\x}_A - \vec{\x}_B \rvert$ 
and of the distances between other invariant sets. This leads to the
asymptotic scaling $N^{5/2}$ for the computation time,  which is not onerous
since the control of one network requires the identification of only one
compensatory perturbation. This should be compared to the $O(\exp(N))$ time that
would be required to determine the basin of attraction  at fixed resolution  by
exhaustive sampling of the state space.
These properties are representative of more general conditions.
 The efficiency and effectiveness of our
approach are expected to be relevant not only for large genetic networks, but
also for large real networks in general,  such as food webs and power grids,
where the characteristic time to failure after the initial perturbation is large
relative to the expected time scales to identify compensatory perturbations.

\begin{figure}[!th]
\centering
\includegraphics[angle=0,width=6.5cm]{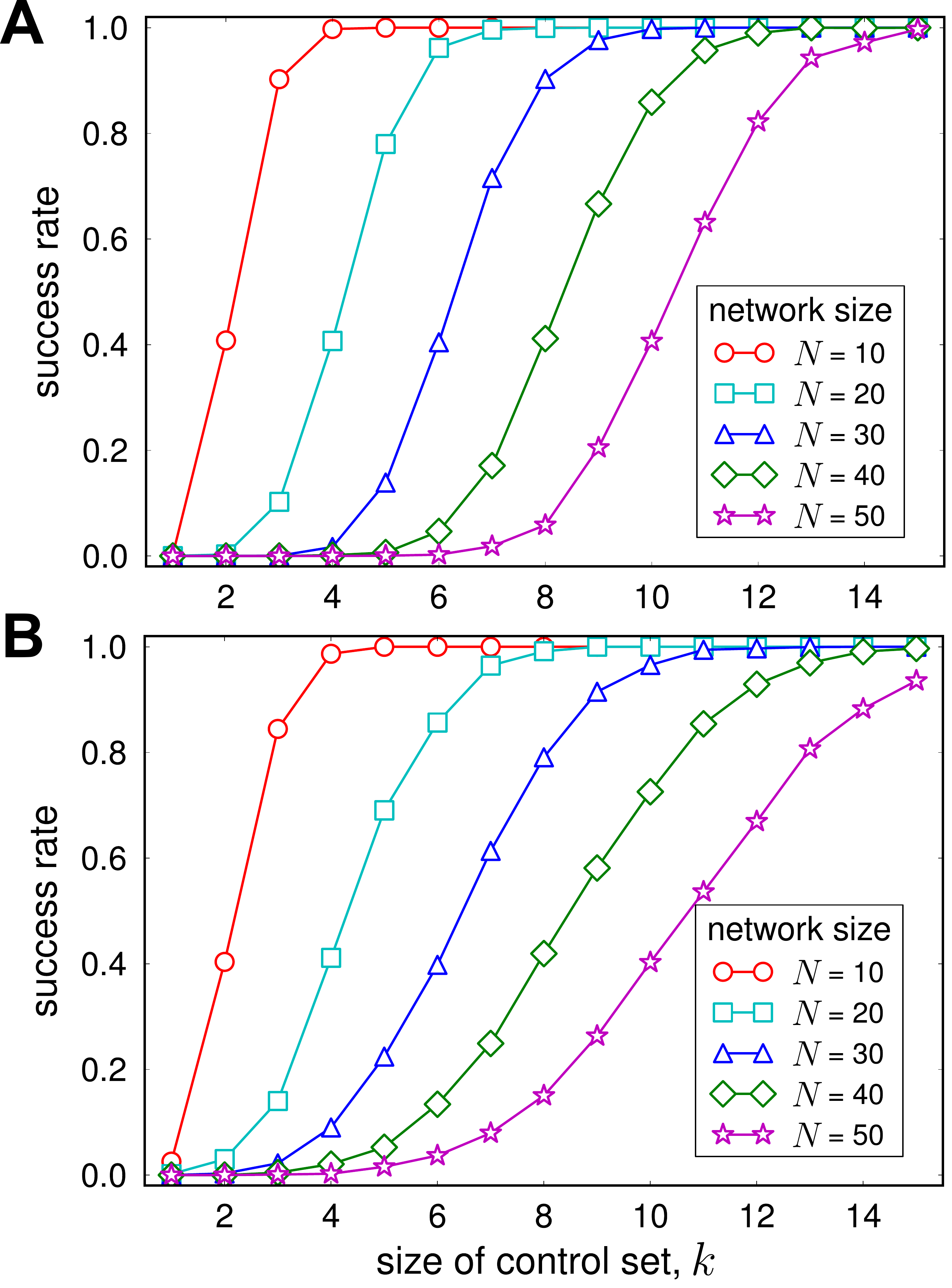}
\caption{\baselineskip 14pt
Control of networks using a small fraction of all nodes. The dynamics
follows (\ref{syncdynamics}) for $\varepsilon=1$ and the control is illustrated
for bringing the network from state $\vec{\x}_B$ to state $\vec{\x}_A$. (A)
Probability that an $N$-node homogeneous network can be controlled by perturbing
only $k$ randomly selected nodes. (B) Corresponding results for heterogeneous
networks with the same average degree (see \emph{Methods}). Each
point represents an average over $1,000$ network realizations for one control
set each. Thus, for control sets consisting of randomly selected nodes, 
heterogeneous networks are  more likely to be controlled than homogeneous
networks if the size of the control set is small.   This  should be contrasted
with the case of control sets formed by the highest-degree nodes (see Fig.\ 6),
where heterogeneous networks are more likely to be controlled for both small and
large control sets.} 
\label{fig4}
\end{figure}

In the preceding analysis, we allowed all nodes in the network to be perturbed,
thus providing an upper bound  for the computational time. In a realistic
situation we may assume that only a subset of the network is made available to
perturbations.  While compensatory perturbations are not expected to exist under
such constraints for very small $\varepsilon$, they become abundant for larger
coupling strengths.  Figure  \ref{fig4} shows the success rate in directing the
system from $\vec{\x}_0 = \vec{\x}_B$ to $\vec{\x}^* = \vec{\x}_A$ for several
network sizes and  a varying number of nodes $k$ in the control set, i.e., the
set of nodes available to be perturbed. For  $\varepsilon = 1$, as considered in
these case,  the success rate increases monotonically with $k$, as expected, but
it is frequently possible to identify an eligible compensatory perturbation
using a small number of nodes in the network. For example, for networks of $50$
nodes,  approximately $40\%$ of all randomly selected $10$-node control sets
will lead to successful control, for both homogeneous (Fig.\ \ref{fig4}A) and
heterogeneous (Fig.\ \ref{fig4}B) networks. Naturally, for the network to be
controlled it suffices to find a single successful control set, indicating that
the networks will often be rescued by a much smaller fraction of nodes. From a
comparison between Fig.\ \ref{fig4}A and Fig.\ \ref{fig4}B it follows that,  for
control sets formed by randomly selected nodes, heterogeneous networks are more
likely to be controlled than homogeneous ones if the size of the control set is
small, while the opposite is true if the control set is large. The analysis
below will allow us to interpret this property of heterogeneous networks as a
consequence of the possibility of having a very high-degree node in small
control sets and a tendency to have a large number of low-degree nodes in large
control sets.  But what distinguishes a successful control set from a set that
fails to control the network? 

\begin{figure}[!th]
\centering
\includegraphics[angle=0,width=16.0cm]{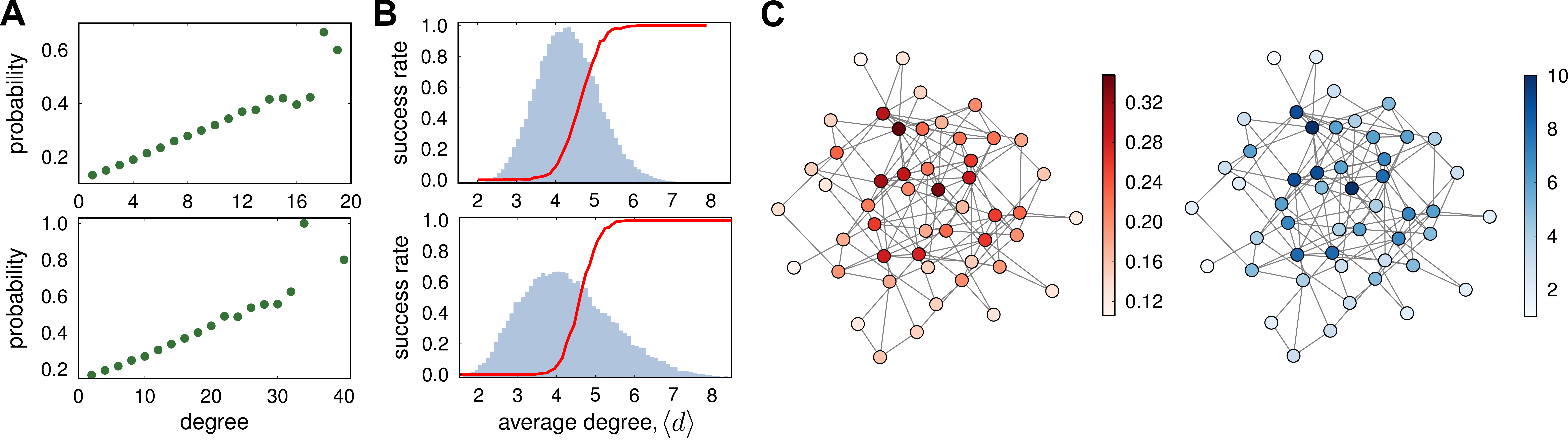}
\caption{\baselineskip 14pt
Enrichment of high-degree nodes in successful control sets. (A) Probability that
a given node of degree $d$ is present in a successful control set for
homogeneous  (top) and heterogeneous (bottom) networks. This probability is
scaled as $P(d) = s(d)/{\cal N}(d)$, where $s$ is the number of times (counting
multiplicity) a node of degree $d$ appears in a successful control set, while
${\cal N}(d)$ is the total number of such nodes appearing in all
successfully-controlled networks. (B)  Dependence of  the success rate on the
average degree of the control set.  Probability (red) for  homogeneous  (top)
and heterogeneous  (bottom)  networks that a random set of $20\%$ of the nodes
can be used to control the corresponding network as a function of the average
degree $\langle d \rangle$ in the set. The background of each panel  (light
blue) shows the distribution of $\langle d \rangle$ for randomly formed control
sets.  (C) Relation between success rate (left) and node degree (right) in an
example network. The color bars show the percentage of successful control sets
involving the given node (red) and the degree of the node (blue). Each point in
panels A and B correspond to an average over $5,000$ network realizations
sampled $10$ times while the statistics in panel C are based on sampling the
given network  $10,000$ times, where $N=50$ and $k=10$ in all cases. The
dynamics, coupling strength, and initial and target states are the same
considered in Fig.\  \ref{fig4}. It follows that the ability to control a
network correlates strongly with the degrees of the nodes in question.
} 
\label{fig5}
\end{figure}

Figure \ref{fig5} shows that the probability of success is strongly determined
by node degree.  This is clear from the probability that a node appears in a
successful control set as a function of its degree $d$ (Fig.\ \ref{fig5}A),
which increases approximately linearly for both homogeneous and heterogeneous
networks.  This is even more evident in the dependence of the success rate on
the average degree $\langle d \rangle$  of the control set  (Fig.\ \ref{fig5}B).
 The success rate exhibits a relatively sharp transition from zero to one as
$\langle d \rangle$ increases, which starts near the peak of the
distribution of average degrees for random control sets (Fig.\ \ref{fig5}B,
background histograms). As shown in Fig.\ \ref{fig5}C  for a homogeneous
network, the success rate of control sets involving a given node correlates
strongly with degree even across relatively small degree differences. This
confirms both that compensatory perturbations can be identified for a large
fraction of control sets and that successful control sets show an enrichment of
high-degree nodes.

\begin{figure}[!th]
\centering
\includegraphics[width=7.0cm]{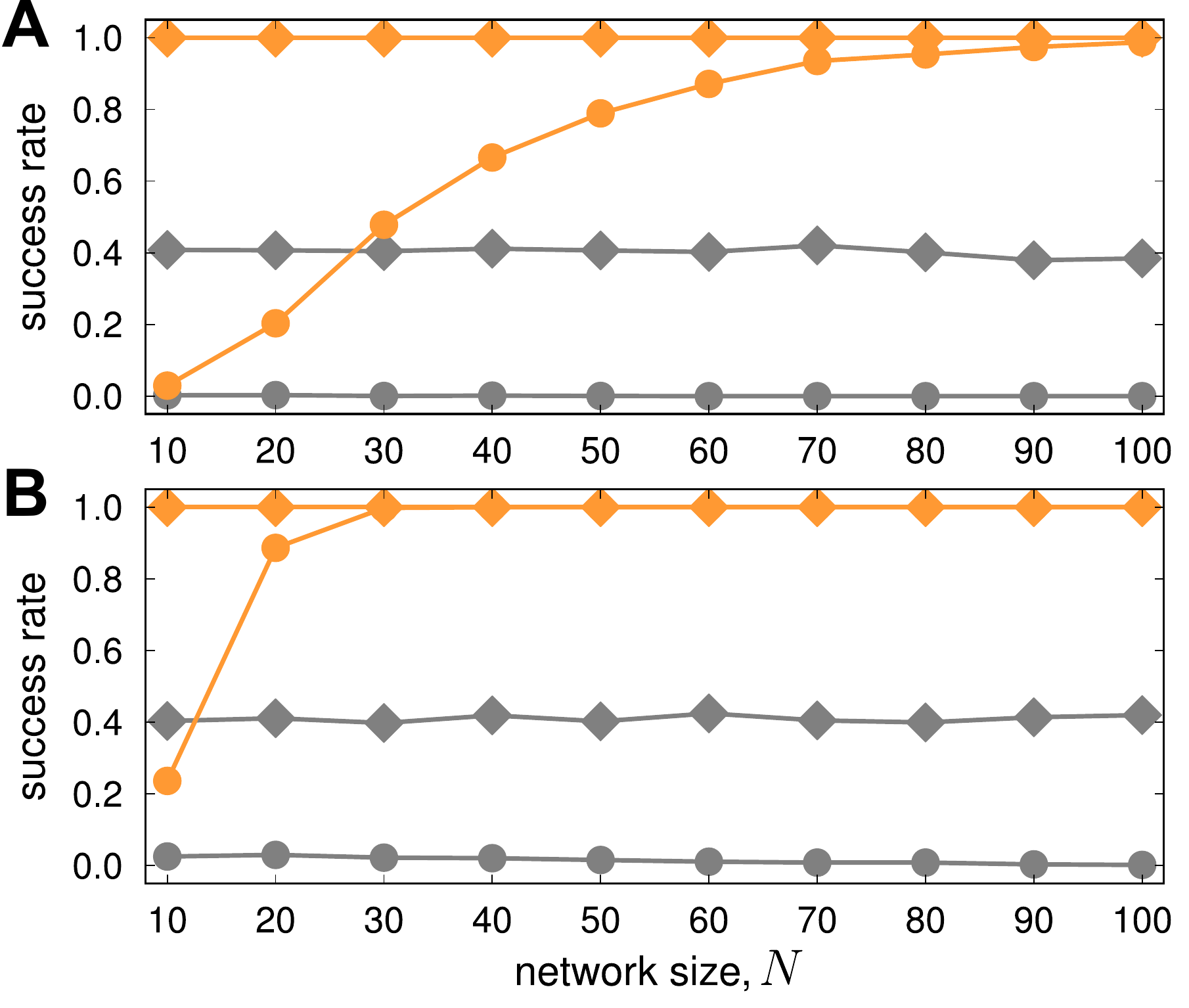}
\caption{\baselineskip 14pt
Increase in rate of successful control through high-degree nodes. 
(A and B)  Probability of successfully controlling homogeneous (A) and heterogeneous (B) networks
with a control set formed by a fraction of randomly selected nodes (gray) or by the same fraction 
of highest-degree nodes (orange).
The fraction of nodes accessible in the control set is selected to be only
$10\%$ (circles) and $20\%$ (diamonds) of the network. Each point represents 
an average over $2,000$ independent network realizations for initial state $\vec{\x}_B$ and target state $\vec{\x}_A$,
for the dynamics and coupling strength considered in Fig.\  \ref{fig4}.
Biasing the intervention towards high-degree nodes dramatically improves the ability to control the network,
permitting success with a small number of nodes where control through randomly selected nodes or low-degree nodes would fail.}
\label{fig6}
\end{figure}

Because successful control sets tend to have a significantly higher fraction of
high-degree nodes, increased success rate can be achieved by biasing the
selection of control-set nodes towards high degrees. This is demonstrated in
Fig.\ \ref{fig6}, where the rate of success increases dramatically when the
control set is formed by the highest-degree nodes in the network.  For example,
for the control sets formed by $20\%$ of the nodes, the success rate approaches
$100\%$  for all network sizes considered, more than doubling when compared to
randomly selected control sets. This increase is more pronounced for more
heterogeneous networks,  as illustrated for control sets limited to $10\%$ of
the nodes (Fig.\ \ref{fig6}), which is expected from Fig.\   \ref{fig5}  and the
availability of higher degree nodes in such networks. This enhanced
controllability of degree-heterogeneous networks for targeted node selection is
important in view of the prevalence of heterogeneous degree distributions in
natural and engineered networks.  We emphasize that  these conclusions follow
from our systematic identification of compensatory perturbations, and that they
cannot be derived from the existing literature.

Note that our approach is fundamentally different from those usually considered
in control theory, both in terms of methods and applicability.  Optimal control 
\cite{Kirk2004}, for example, is based on identifying an admissible
(time-dependent) control ${\mathbf u}(t)$ such that the modified system
$d{\mathbf x}/{dt}={\mathbf F} + {\mathbf u}$ will optimize a given cost
function.  Control of chaos \cite{ott95}, which is used to convert an otherwise
chaotic trajectory into a periodic one, is based on the continuous application
of unconstrained small time-dependent perturbations to align the stable manifold
of an unstable periodic orbit with the trajectory of the system. Approach to the
relevant orbit can be facilitated by the method of targeting \cite{bollt2003},
which, like control  of chaos itself,  can only be applied to move within the
same ergodic component. Here, in contrast, we bring the system to a different
component, to a state that is already stable, and we do so using one (or few)
finite-size perturbations, which are  forecast-based rather than feedback-based.
 Our approach is suited to a wide range of processes in complex networks, which
are often amenable to modeling but offer limited access to interventions, making
the use of punctual interventions that benefit from the natural stability of the
system far more desirable than other conceivable alternatives. \\ \\

\noindent {\Large\bf Discussion} \\
\medskip

The dynamics of large natural and man-made networks are usually highly
nonlinear, making them complex not only with respect to their structure but
also with respect to their dynamics. This is particularly manifest when the
networks are brought away from equilibrium, as in the event of a strong
perturbation. Nonlinearity has been the main obstacle to the control of such
systems. Progress has been made in the development of algorithms for
decentralized communication and coordination  \cite{bullo2009}, in the
manipulation of two-state Boolean networks \cite{Shmulevich2009},  in network
queue control  problems \cite{Meyn2008}, and other complementary areas. Yet, the
control of far-from-equilibrium networks with self-sustained dynamics, such as
metabolic networks, power grids, and food webs, has remained largely
unaddressed. Methods have been developed for the control of networks
hypothetically governed by linear dynamics \cite{lin_ieee_1974}. But although
linear dynamics may approximate an orbit locally, it does not permit the
existence of different stable states observed in real networks and does not
account for basins of attraction and other global properties of the state space.
These global properties are crucial because they underlie network failures and,
as shown here, also provide a mechanism for the control of numerous networks. 

The possibility of directing a complex network to a predefined dynamical state
offers an unprecedented opportunity to harness such highly structured systems.
We have shown that this can be achieved under rather general conditions  by
systematically designing compensatory perturbations that, without requiring the
computationally-prohibiting explicit identification of the basin boundaries, 
take advantage of the full basin of attraction of the desired state, thus
capitalizing on (rather than being  obstructed by) the nonlinear nature of the
dynamics.  While our approach makes use of constrained optimization
\cite{Bazaraa2006}, the question at hand cannot be formulated as a simple
optimization problem in terms of an aggregated objective function, such as the
number of active nodes. Maximizing this number by ordinary means can lead to
local minima or  transient solutions that then fall back to asymptotic states
with larger number of inactive nodes.  {\it A priori} identification of the
stable state that enjoys the desired properties is thus an important step in our
formulation of the problem.

Applications of our  framework show that the approach is effective even when
compensatory perturbations are limited to a small subset of all nodes in the
network, and when constraints forbid bringing the network directly to the target
state.   This is of outmost importance for applications because, in large
networks,  perturbed states in which the desired state is outside the range of
admissible perturbations is expected  to be the rule rather than the exception.
From a network perspective,  this leads to counterintuitive situations in which
the compensatory perturbations oppose the direction of the target state and
consist, for example, of suppressing nodes whose activities are already
diminished relative to the target, but that lead the system to eventually evolve
towards that target. These results are surprising in light of the usual
interpretation that nodes represent ``resources" of the network, to which we
intentionally (albeit temporarily) inflict damage with a compensatory
perturbation.  From the state space perspective,  the reason for the existence
of such locally deleterious perturbations that have globally beneficial effects
is that the target's basin of attraction, being nonlocal, can extend to the region of
feasible perturbations even when the target itself does not.

Our results based on randomly-generated networks also show that control is far
more effective when focused on high-degree nodes. The effect is more pronounced
in more heterogenous networks, making it particularly relevant for applications
to real systems, which are known to often exhibit fat-tailed degree
distributions \cite{barabasi_science_1999,Caldarelli_book_2007}.   More
generally, we posit that the rate of successful control on a network will be
strongly biased  towards nodes with high centrality,  including degree
centrality, as shown here, and possibly other measures of centrality in the case
of networks that deviate from random, such as closeness and betweenness.  This
suggests that in the situation where only a limited intervention is possible,
one should focus primarily on central nodes in order to direct the entire
network to a desired new state.  We have demonstrated, moreover, that even when
all nodes in the network are perturbed, the identification of compensatory
perturbations is computationally inexpensive and scales well with the size of
the network, allowing the analysis of very large networks.

We have motivated our problem assuming that the network is away from its
desirable equilibrium due to an external perturbation. In that context, our
approach provides a methodology for real-time rescue of the network, bringing it
to a desirable state before it reaches a state that can be temporarily or
permanently irreversible, such as in the case of cascades of overload failures
in power grids or cascades of extinctions in food webs, respectively.  We
suggest that this can be important for the conservation of ecological systems
and for the creation of self-healing infrastructure systems. More generally, as
illustrated in our genetic network examples, our approach also applies to change
the state of the network from one stable state to another stable state, thus
providing a mechanism for ``network reprogramming". 

As a context to interpret the significance of this application, consider the
reprogramming of differentiated (somatic) cells from a given tissue to
pluripotent stem cell state, which can then differentiate into cells of a
different type of tissue.  The seminal experiments demonstrating this
possibility involved continuous overexpression of specific genes
\cite{Takahashi_cell_2006}, which is conceivable even under the hypothesis that
cell differentiation is governed by the loss of stability of the stem cell state
\cite{ Huang_DevBiol_2007}. However, the recent demonstration that the same can
be achieved by temporary expression of few proteins \cite{kim_stem_2009}
indicates that  the stem cell state may have remained stable (or metastable)
after differentiation,  allowing interpretation of this process in the context
of the interventions considered here. This is relevant because a very small
fraction of a population of treated cells is found to undergo the transition
back to the stem cell state,  suggesting that much could be learned from a
modeling approach capable of identifying the perturbations most likely to direct
the cellular network to adopt a new stable state. While induced pluripotency is
an example par excellence of network reprograming, the same concept extends far
beyond this particular system.  Taken together, our results provide foundation
for the control and rescue of network dynamics, and as such are expected to have
implications for the development of smart traffic and power-grid networks, of
ecosystems and Internet management strategies, and of new interventions to
control the fate of living cells. \\ \\

\noindent {\Large\bf Methods} \\
\medskip

\noindent {\bf Constraints on incremental perturbations.}\  
The iterative procedure behaves
well as long as the linearization involved in Eq.\ (\ref{variational}) remains
valid at each step. The incremental perturbation at the point of closest
approach, $\delta \x(t_c)$, selected under constraints (\ref{constraints}) alone
will generally have a nonzero component along a stable subspace of the orbit
$\x(t)$, which will result in  $\delta \x_0$ larger than $\delta \x(t_c)$ by a
factor $O(\exp|\lambda_s^{t_c}(t_c-t_0)|)$, where $\lambda_s^{t_c}$ is the
finite-time Lyapunov exponent of the eigendirection corresponding to the
smallest-magnitude eigenvalue of ${\bf M}(\x_0, t_c)$. In a naive implementation
of this algorithm, to keep $\delta \x_0$ small for the linear transformation to
be valid, the size of $\delta \x(t_c)$ would be negligible, leading to
negligible progress. This problem is avoided by optimizing the choice of 
$\delta \x(t_c)$ under the constraint that the size of $\delta \x_0$ is bounded above.
Another potential problem is when the perturbation causes the orbit to cross an
intermediate  basin boundary before reaching the final basin of attraction. All
such events can be detected by monitoring the difference between the linear
approximation and the full numerical integration of the orbit, without requiring
any prior information about the basin  boundaries. Boundary crossing is actually
not a problem because the closest approach point is reset to the new side of the
basin boundary.   To assure that the method will continue to make progress, we
solve the optimization problem under the constraint that the size of 
$\delta \x_0$ is also bounded below, which means that we accept increments $\delta \x_0$
that may temporarily increase the distance from the target (to avoid
back-and-forth oscillations, we also require the inner product between two
consecutive increments $\delta \x_0$ to be positive).   These upper and lower
bounds can be expressed as
\begin{equation}
\epsilon_0\le |\delta \x_0| \le \epsilon_1.
\label{constraints2}
\end{equation}
This can lead to $|\delta \x(t_c)|\gg\epsilon_1$ due to components along the
unstable subspace, but in such cases the vectors can be rescaled after the
optimization.  At each iteration, the problem of identifying a perturbation
$\delta \x_0$ that incrementally moves the orbit toward the target under
constraints (\ref{constraints}) and (\ref{constraints2}) is then solved as a
constrained optimization problem.\\

\noindent {\bf Nonlinear optimization.}\ 
The optimization step of the iterative control
procedure consists of finding the small perturbation $\delta \x_0$ that
minimizes the remaining distance between the target, $\target$, and the system
orbit $\x(t)$ at its time of closest approach, $t_c$. Constraints are used to
both define the admissible perturbations (\ref{constraints}) and also, as
described in the previous paragraph, limit the magnitude of $\delta \x_0$
(\ref{constraints2}). The optimization problem to identify $ \delta \x_0$ can
then be  succinctly written as:
\begin{align} \label{nlp}
\textrm{min} & & \lvert \target - (\x(t_c) + {\bf M}(\x_0', t_c) \cdot \delta \x_0) \rvert &\\
\textrm{s.t.} & & \mathbf{g}(\x_0, \x_0' + \delta \x_0) & \leq 0 \label{ieq_constraint} \\
& & \mathbf{h}(\x_0, \x_0' + \delta \x_0) & =  0 \label{eq_constraint} \\
& & \eps_0 \; \leq \lvert \delta \x_0 \rvert \; & \leq \eps_1 \label{size_constraint} \\
& & \delta \x_0 \cdot \delta \x_0^{p} & \geq 0, \label{direction_constraint}
\end{align}
where (\ref{direction_constraint}) is enforced starting from the second
iteration,  and $\delta \x_0^{p}$ denotes the incremental perturbation from the
previous iteration. Formally, this is a nonlinear programming (NLP) problem, the
solution of which is complicated by the nonconvexity of the constraint
(\ref{size_constraint}) (and possibly (\ref{ieq_constraint}) and
(\ref{eq_constraint})). Nonetheless, a number of algorithms have been developed
for the efficient solution of NLP problems, among them Sequential Quadratic
Programming (SQP) \cite{sqp}. This approach solves (\ref{nlp}) as the limit of a
sequence of quadratic programming subproblems, in which the constraints are
linearized in each sub-step. For all calculations, we used the SQP algorithm
\cite{kraft} implemented in the SciPy scientific programming package
(http://www.scipy.org/). Note that this implementation does not require
inverting matrix ${\bf M}$ (cf. Eq.\ (\ref{variational})).\\

\noindent {\bf Comparison with backward integration.}\
Our approach should be compared with an apparently simpler alternative. Rather
than keeping track of the  variational matrix ${\bf M}(\x_0,t)$, which requires
the integration of $n^2$ additional  differential equations at every iteration,
one could imagine using backwards integration of a trajectory starting at
$\x(t_c)+\delta \x(t_c)$ to identify a suitable initial  perturbation.  This
alternative procedure suffers the critical drawback that, for a particular
choice of $\delta \x(t_c)$ (magnitude and direction), it is not certain that the
time-reversed orbit will ever strike the feasible region defined by
(\ref{constraints}). This is particularly so in realistic situations where only
a fraction of the nodes are accessible to perturbations, resulting in a feasible
region of measure zero in the full $n$-dimensional state space. \\

\noindent {\bf Termination criteria and parameters.}\
In all simulations,  the control procedure is terminated if the updated initial
condition attracts to within $\kappa = 0.01$ of the target state within $\tau
=10,000$ time units.  Otherwise,  we terminate the search if a compensatory
perturbation is not found after a fixed number of iterations, which we set to be
$I = 1,000$. In general this number should be of the order of $L$/$\eps_0$,
where $L$ is the characteristic linear size of the feasible region. For each
iteration, we use $T = 10$ time units within the  integration step that
identifies $t_c$, which was estimated based on the time to approach the
undesirable stable state. In the examples of Fig.\ \ref{fig2}, we used the
parameters $\eps_0 = 0.001$ and $\eps_1 = 0.01$ for the optimization step. To
lighten the computational burden in our Monte Carlo analysis, we used a more
relaxed choice of $\eps_0 = 0.005$ and $\eps_1 = 0.05$ for the genetic networks 
presented in Figs.\ \ref{fig3}--\ref{fig6}.\\

\noindent {\bf Construction of genetic networks.}\ 
The networks are grown starting with a
$d$-node connected seed network, by iteratively attaching a new node and
connecting this node to each pre-existing node $i$  with probability  $d\times P_i$,
where $\sum_i P_i=1$. The connections are assumed to be unweighted and
undirected. We reject any iteration resulting  in a degree-zero node, thereby
ensuring that the final $N$-node network is  connected and  has average degree
$\ge 2d$ for large $N$.  Networks are generated for both uniform attachment
probability, where $P_i= 1/N^{s}$ and $N^{s}$ is the number of nodes in the
network at iteration $s$, and for linear preferential attachment, where  
$P_i= k^{s}_i/\sum_j k^{s}_j$ and $k^{s}_i$ is the degree of node $i$ at iteration
$s$. The former leads to networks with asymptotically exponential distribution
of degrees, which we refer to as homogeneous networks, while the latter leads to
networks with asymptotically power-law distribution of degrees, which we refer
to as heterogeneous networks \cite{barabasi_science_1999}. By construction, the
resulting average degree is essentially the same for both homogeneous and
heterogeneous networks. In our simulations, we focused on networks with $d=2$.\\

\noindent {\bf Acknowledgments}
\medskip

\noindent
The authors thank Jie Sun for illuminating discussions and the Brockmann Group
for use of their computer cluster.  This work was supported by the  National
Science Foundation under Grant DMS-1057128 and the National  Cancer Institute
under Grant 1U54CA143869-01.

\newpage

\newpage

\newpage

\end{document}